\documentclass[aps,groupedaddress,fleqn,twocolumn,longbibliography]{revtex4-2}
\usepackage{graphicx}
\usepackage{xcolor}
\usepackage{amsmath,amssymb}
\usepackage{float}
\usepackage{physics}
\usepackage{amsfonts}
\usepackage{verbatim}
\usepackage{flushend}
\usepackage{balance}
\usepackage{endnotes}
\usepackage{footnote}
\usepackage{adjustbox}
\usepackage{gensymb}
\usepackage{subfigure}
\usepackage{float}
\usepackage{epigraph}
\usepackage{xcolor}
\usepackage[utf8]{inputenc}
\usepackage{setspace}
\newcommand{\beq}{\begin{eqnarray}}
\newcommand{\eeq}{\end{eqnarray}}
\newcommand{\kb}{k_{\mathrm{B}}}
\newcommand{\av}{\alpha_V}
\newcommand{\ath}{\alpha_T}
\usepackage{amsmath}

\usepackage[normalem]{ulem}

\begin{document}

\title{Thermal conductivity and thermal diffusivity of molten salts: insights from molecular dynamics simulations and fundamental bounds}

\author{ C. Cockrell$^{1,2,*}$, M.Withington $^{2}$, H. L. Devereux$^{2}$,  A. M. Elena$^{3}$, I. T. Todorov$^{3}$, Z. K. Liu${^4}$, S. L. Shang${^4}$, J. S. McCloy$^{5}$, P. A. Bingham${^6}$, K. Trachenko ${^2}$}
\address{$^1$ Nuclear Futures Institute, Bangor University, Bangor, LL57 1UT, UK}
\address{$^2$ School of Physical and Chemical Sciences, Queen Mary University of London, Mile End Road, London, E1 4NS, UK}
\address{$^3$ Scientific Computing Department, Science and Technology Facilities Council, Daresbury Laboratory, Keckwick Lane, Daresbury, WA4 4AD, UK}
\address{$^4$ Department of Materials Science and Engineering, The Pennsylvania State University, University Park, PA 16802, USA}
\address{$^5$ School of Mechanical and Materials Engineering, Washington State University, Pullman, WA, USA}
\address{$^6$ Materials and Engineering Research Institute, Sheffield Hallam University, Sheffield, S1 1WB, UK}
\address{* Corresponding author, e-mail: c.cockrell@bangor.ac.uk}

\begin{abstract}
We use extensive molecular dynamics simulations to calculate thermal conductivity and thermal diffusivity in three common molten salts, LiF, LiCl, and KCl. Our analysis includes the total thermal conductivity and intrinsic conductivity, excluding mass currents, measured experimentally. The latter shows good qualitative agreement with the experimental data. We also calculate their key thermodynamic properties such as constant-pressure and constant-volume specific heats. We subsequently compare the results to the lower bound for thermal diffusivity expressed in terms of fundamental physical constants. Using this comparison and recent theoretical insights into thermodynamic and transport properties in liquids, we interpret thermal properties on the basis of atomistic dynamics and phonon excitations. We finally find that thermal diffusivity of molten salts is close to their kinematic viscosity.
\end{abstract}

\maketitle

\section{Introduction}

Molten salts are used in a number of important applications including energy generation. This includes nuclear energy where coolants are required to transfer heat to generate electricity. Water is the primary coolant in current nuclear reactors, however, it has its disadvantages, such as high-reactivity \cite{US_nuclear2011}. Traditional coolants cannot be recycled after use and contribute to a large proportion of nuclear waste. One motivator for the development of molten salt reactors is they allow for a recyclable coolant which would dramatically reduce the amount of waste in nuclear energy production. Fuel rods make up 3\% of the nuclear waste which needs to be disposed. There are currently some liquids being considered as a fuel transport fluid and/or coolant which could be processed and recycled in order to minimise nuclear waste. These liquids include molten salts which are the subject of this study \cite{Rosenthal1970,FRANDSEN2020,Allen2007,Tang2015}.

Flow and transfer of heat are two primary functions of molten salts in these applications \cite{zhao-review}. Thermal conductivity and thermal diffusivity are important characteristics of these liquids to consider. These properties can vary drastically depending on the operating temperature and pressure and are responsible for the flow, diffusion, and thermal transport properties of the liquids. This makes it important to collect thermal data and transport across a wide range of temperatures and pressures and to understand physical mechanisms setting thermal conductivity of molten salts.

As recently reviewed \cite{zhao-review}, acquiring reliable thermal conductivity data is difficult. First, high temperature experiments are challenging due to corrosion and obfuscating effects of convection and radiation. The experimental thermal conductivity in different studies can differ by a factor of 3, substantially affecting the projected heat output in energy production and hence the designs of solar and nuclear power generators. Second, commonly used theories such as kinetic gas theories do not describe the data well \cite{zhao-review}. Hence improved understanding from theory and modelling is needed.

Thermal conductivity $\kappa$ and thermal diffusivity $\alpha_T$ are strongly dependent on temperature, pressure and system structure. Despite this sensitivity, it was proposed that thermal diffusivity has a lower bound, $\alpha_m$, set by fundamental physical constants as 

\begin{equation}
\alpha_m=\frac{1}{4\pi}\frac{\hbar}{\sqrt{m_e m}}
\label{alphamin}
\end{equation} 

\noindent where $\hbar$ is Planck's constant, $m_e$ is the mass of electron and $m$ the mass of the particle (or average mass for multi-component systems) \cite{prbthermal,myreview,brareview}.

The scale of $m$ is set by the proton mass $m_p$, giving rise to fundamental thermal diffusivity 

\begin{equation}
\alpha_f=\frac{1}{4\pi}\frac{\hbar}{\sqrt{m_e m_p}} \approx 10^{-7} ~\frac{{\rm m}^2}{{\rm s}}
\label{alphafund}
\end{equation} 

In addition to applications where high thermal conductivity is desirable, exploring the theoretical minimum of thermal diffusivity has the benefit of setting a goal in any optimisation of thermal insulation products where the low values of $\alpha_T$ are sought.

The lower bound \eqref{alphamin} agrees with experimental data in noble, molecular and hydrogen-bonded liquids \cite{prbthermal}. However, no comparison to ionic liquids has been made.

It is useful at this point to explain the origin of the minimum of thermal diffusivity. For systems in which the heat is carried by phonons, such as insulating solids and liquids, thermal conductivity $\kappa$ is \cite{ashcroft}

\begin{equation}
\kappa = cvl
\label{kappa}
\end{equation}
\noindent 

\noindent where $c$ is the heat capacity per unit volume, $v$ the speed of sound, and $l$ the phonon mean free path. The thermal diffusivity $\alpha_T=\frac{\kappa}{\rho c_P}$ ($c_P$ is the specific isobaric heat capacity) is then

\begin{equation}
    \alpha_T = vl
    \label{alphat}
\end{equation}

At lower temperatures, liquid dynamics comprise ``solid-like" particle oscillations and occasional diffusive jumps. \textcolor{black}{This means that liquids exhibit a collective excitation (phonon) spectrum similar to those sustained by solids \cite{ropp}, except that transverse phonons do not exist at long wavelengths \cite{ropp}. These spectra can be discerned using inelastic scattering experiments and the long wavelength ``gap" was recently observed in several liquid metals \cite{inui}.} In liquids, therefore, heat is mostly carried by phonons and the phonon mean free path $l$ decreases with temperature increase. These liquid phonons are propagating collective modes which contribute to the thermal properties of liquids in much the same way that formal crystalline phonons do in solids \cite{ropp}. Together with the decrease of the speed of sound $v$ with temperature, Eq. \eqref{alphat} implies that $\alpha_T$ decreases with temperature. This is in contrast to gaslike dynamics where particles move in straight lines between collisions. Here $v$ is proportional to the thermal velocity of the molecules and $l$ corresponds to the the molecular mean free path instead of the phonon mean free path. Both of these quantities increase with temperature. The minimum of thermal diffusivity $\alpha_T$ appears as a result of the crossover between liquidlike and gaslike dynamics \cite{prbthermal}.

In this paper, we use extensive molecular dynamics (MD) simulations to calculate thermal conductivity and thermal diffusivity in three common molten salts, LiF, LiCl, and KCl. Our calculations include the total thermal conductivity and intrinsic conductivity, the latter of which excludes the heat advected by partial mass currents and corresponds more closely with experimental measurements. By exploring different phase diagram paths and incorporating many more state points than previous studies \cite{Ohtori2009,Galamba2007}, we examine the response to density and temperature of each type of thermal conductivity, and demonstrate that thermal conductivity meaningfully evolves with temperature at constant density, contrary to previous predictions based on sparser data \cite{Ohtori2009} We also calculate their thermodynamic properties such as isobaric and isochoric specific heats. We use the lower bound \eqref{alphamin} as a theoretical benchmark to discuss our modelling results and interpret them and find that the calculated thermal diffusivities conform to the theoretical bound. Together with earlier theories of thermodynamic and transport properties in liquids, we interpret the calculated thermal properties of molten salts and their evolution with temperature and composition on the basis of atomistic dynamics and phonon excitations in liquids. We finally find that thermal diffusivity of molten salts is close to another transport property, namely their kinematic viscosity. We discuss the origin and implications of this closeness.


\section{Mechanisms of thermal conductivity and simulations methods and details}
All simulations were performed with the DL\_POLY MD package \cite{Todorov2006}. For LiF, we used an empirical potential \textcolor{black}{derived and employed in previous work \cite{withington2024,Kanian2009,Luo2016}. The potential consists of electrostatic interactions and a Buckingham term:}
\textcolor{black}{\begin{equation}
    \label{eqn:buck}
    \phi(r_{ij}) = \frac{1}{4 \pi \epsilon_0}\frac{q_i q_j}{r_{ij}} + A_{ij} \exp\left(-B_{ij} r_{ij}\right) - \frac{C_{ij}}{{r_{ij}}^6},
\end{equation}}
\textcolor{black}{with ions being assigned their formal $\pm 1$ charges. Potential parameters are reported in Tab. \ref{tab:params}.} For LiCl and KCl we used a Born-Huggins-Mayer potential derived  using the Fumi and Tosi rules as outlined in Sangster's paper \cite{Sangster1976}. \textcolor{black}{These potentials run the risk of exhibiting pathological behaviour, due to the attractive terms dominating the repulsive term at small separations. If an atom surmounts the repulsive maximum from its neighbour, the two will become unphysically bound. This is vanishingly likely near the melting point, however it can be disruptive at higher temperatures, particularly along isochores where the density is fixed at a high value. Avoiding this problem informed the maximum temperatures we examine here.}

\textcolor{black}{In each of the systems simulated here, the Coulomb term dominates the potential energy. From melting point to boiling point, in each composition, the electrostatic potential energy constitutes 80-90\% of the total potential energy. This domination of the electrostatic energy is seen by the near-equality of like-ion (\textit{e.g.} Li-Li, F-F) partial radial distribution functions in the molten state \cite{McGreevy1987}. The dynamics of liquids interacting \textit{via} such potentials are expected to be driven by both charge and mass fluctuations, corresponding roughly to acoustic and optical modes in a solid's phonon spectrum. The potentials do not allow for the polarisation of any ions, which introduces some discrepancy between simulated and experimental data. However, the structure \cite{Copley1976,Copley1976b}, and dynamics \cite{Caccamo1980} of these models are qualitatively similar to, though sometimes quantitatively somewhat different from, those inferred from experimental data. Furthermore, the collective dynamical features of molten salts, as discussed in the Introduction, are understood from simulations to be qualitatively the same as those of ``simple" liquids (which do not feature free charges) \cite{Adams1977}. The presence of these modes and their contribution to the total system energy is well established from neutron scattering experiments \cite{McGreevy1987}. The use of non-polarisable potentials is therefore a valuable source of information in the relationship between the dynamics and thermodynamics of ionic liquids and molten salts in general.}

\begin{table*}[ht]
  \centering
  \begin{tabular}{ |c|c|c|c|c|c|c| } 
  \hline
Species $i$ & Species $j$ & $A_{ij}$ (eV) & $B_{ij}$ (\AA$^{-1}$)  & $C_{ij}$ (eV \AA${^6}$) \\
 \hline
    Li & Li  & 98.9388 & 3.34 & 0.045568 \\ 
    Li & F & 229.026 & 3.34 & 0.499376 \\
    F & F & 420.537 & 3.34 & 0.905119\\
\hline
\end{tabular}
\caption{Parameters modelling pair interactions in LiF.}
\label{tab:params}
  \end{table*}
  

We first equilibrated the systems of 800 atoms at each target pressure and temperature in the constant-pressure and constant-temperature ensemble (NPT) using \textcolor{black}{the Hoover thermostat and barostat} for 10$^6$ time steps with a fixed simulation time step of 0.001 ps. We then ran a constant energy and volume ensemble (NVE) where no data was collected to act as a buffer period due to the change in ensembles. Then we performed production runs for 10$^6$ steps in NVE where the thermal conductivity data was collected. To calculate the isobaric heat capacity we performed another 10$^6$ step production run in the NPT ensemble.  It has been demonstrated that viscosity can be calculated accurately using small system sizes \cite{Kang-Sahn2019,withington2024,Cockrell2021}. In this study, we calculated the thermal conductivity of systems with variables sizes, between 500 and 6400 atoms, finding no size dependence in this range.

We calculate thermal conductivity using the Green-Kubo method as \cite{Balucani1995,Zwanzig1965}

\begin{equation}
\label{eqn:kappa}
    \kappa \;=\; \frac{V}{3\kb T^2} \int_0^\infty \dd t \ \langle \mathbf{q}(0) \cdot \mathbf{q}(t)\rangle, 
\end{equation}

\noindent with $T$ the (mean) temperature and $V$ the system volume. Here, $\mathbf{q}(t)$ is the microscopic energy current density at time $t$, with atomistic definition:

\begin{equation}
    \label{eqn:qcurrent}
    \mathbf{q}(t) = \frac{1}{V} \sum_{i=1}^N \biggl(u_i(t) \dv{\mathbf{r}_i(t)}{t} + \frac{1}{2} \sum_{j\neq i}\mathbf{f}_{ij}(t)\cdot \dv{\mathbf{r}_i(t)}{t} \mathbf{r}_{ij}(t)\biggr),
\end{equation}
with $u_i$ the total (kinetic plus potential) energy of particle $i\in 1, 2, \ldots N$, $\mathbf{f}_{ij}$ the total force impressed upon particle $i$ by particle $j$, $\mathbf{r}_{ij}$ the inter particle separation vector from particle $i$ to particle $j$, and $\mathbf{r}_i$ is the position of particle $i$.

We note that Eq. \eqref{eqn:kappa} is not the only possible definition of a thermal conductivity in multicomponent system. In particular, this definition is a fluctuation-dissipation relation for the molecular heat current, without necessarily discriminating that the heat flow comes from a temperature gradient alone. The formal definition of thermal conductivity $\lambda$ directly relates the heat flow $\mathbf{J}^{q}$ to the temperature gradient:
\begin{equation}
\label{eqn:fourier}
    \mathbf{J}^q = -\lambda \ \mathrm{grad}(T).
\end{equation}

The freedom of atomic species to flow and diffuse with reference to the barycentric velocity (which vanishes in MD), thereby advecting enthalpy, means that thermal conductivity inferred from experiments with vanishing partial momentum currents will be smaller than the  thermal conductivity as calculated in Eq. \eqref{eqn:kappa} \cite{Armstrong2014}, with the discrepancy increasing when mass (or charge) ratios increase. We will refer to $\lambda$ as the intrinsic thermal conductivity, as it explicitly excludes enthalpy advection, and $\kappa$ as the total thermal conductivity, as it implicitly includes it. It is nonetheless instructive to examine the total thermal conductivity $\kappa$ in MD as it directly incorporates all irreversible phenomena contributing to heat flow.  The mechanisms motivating a minimal thermal diffusivity, which depend on fundamental energy and mass scales, apply more directly to the intrinsic thermal conductivity $\lambda$, as mass diffusion (occurring \textit{via} atomic jumps in the liquid) is a thermal activation effect. The intrinsic thermal conductivity, $\lambda$ is calculated in MD by explicitly subtracting mass transfer effects according to \cite{Armstrong2014}:
\begin{equation}
    \label{eqn:lambda}
    \lambda = \frac{V}{3 \kb T^2} \left( L_{q,q} - \frac{L_{q,1}^2}{L_{1,1}} \right),
\end{equation}
where the phenomenological coefficients are, as in the single component case, Green-Kubo integrals:
\begin{equation}
    L_{1,1} = \int_0^\infty \dd t \ \langle \mathbf{j}_1(0) \cdot \mathbf{j}_1(t)\rangle,
\end{equation}
\begin{equation}
    L_{q,1} = \int_0^\infty \dd t \ \langle \mathbf{q}(0) \cdot \mathbf{j}_1(t)\rangle,
\end{equation}
\begin{equation}
    L_{q,q} = \int_0^\infty \dd t \ \langle \mathbf{q}(0) \cdot \mathbf{q}(t)\rangle.
\end{equation}
Here $\mathbf{j}_1(t)$ is the partial momentum density of component 1:
\begin{equation}
    \label{eqn:masscurrent}
    \mathbf{j}_1(t) = \frac{1}{V} \sum_{i}^{\{ 1 \}} m_i \dv{\mathbf{r}_{i}(t)}{t},
\end{equation}
with the sum performed only over atoms of species 1. Since the total momentum of an MD cell is fixed at 0, $\mathbf{j}_2(t) = - \mathbf{j}_1(t)$ in two-component systems. For the same reason, the momentum current vanishes in the single-component case, and, macroscopically, the energy is conveyed by the centre of mass and therefore considered as convected rather than as a heat flow, resulting in $\kappa = \lambda$. \textcolor{black}{The maximum correlation time taken in approximating these integrals is 5 ps, which is more than enough time for the integrand to decay to noise around 0. Using very large correlation times (especially in lieu of using multiple initial conditions) is undesirable, as integrating the tail fluctuating around zero amounts to integrating a random walk, whose variance grows with time \cite{Zhang2015}}

The thermal diffusivity, $\alpha_T$, is the coefficient relating the temporal evolution of temperature to its spatial distributtion in a system and has the same dimension as the diffusion coefficient. Its definition is:
\begin{equation}
\label{eqn:thdif}
    \alpha_T \;=\;\frac{\lambda}{\rho \ c_P},
\end{equation}
where $\rho$ is the mass density.


The isobaric heat capacity, $C_P$ in the NPT ensemble can be calculated directly using fluctuations in enthalpy, $H$, according to:
\begin{equation}
\label{eqn:cp}
    C_P\;=\;\frac{\langle H^2 \rangle - \langle H \rangle ^2}{\kb T^2}.
\end{equation}

We find that this expression converges slowly in molecular dynamics simulations. \textcolor{black}{Specifically, though most data presented here are collected by averaging 20 individual trajectories each lasting 1 ns, on some limited temperature and pressure points we tested up to 80 initial conditions of length 4 ns. The error in the calculated $C_P$, though it did decrease, did not do so by very much. The heat capacity in the NPT ensemble is therefore very trajectory dependent.} We therefore calculate $C_P$ from the isochoric heat capacity, $C_V$ using the relation

\begin{equation}
    \label{eqn:heatcapacityrelation}
    C_P = C_V + V T \av^2 B,
\end{equation}

\noindent where $\av$ is the isobaric coefficient of thermal expansion and $B$ is the isothermal bulk modulus. These quantities take the following forms in the NPT ensemble:

\begin{equation}
    \label{eqn:alpha}
    \av = \frac{1}{\kb T^2} \frac{\langle H V \rangle - \langle H \rangle \langle V \rangle}{\langle V \rangle},
\end{equation}

and
\begin{equation}
    \label{eqn:bulk}
    B = \frac{\langle V \rangle \kb T}{\langle V^2 \rangle - \langle V \rangle^2},
\end{equation}
with $V$ the system volume. \textcolor{black}{The bulk modulus converges very quickly, and the thermal expansion coefficient does so acceptably, suggesting that the sampling of enthalpy in the NPT ensemble is the culprit.} The fastest-converging method of calculating $C_V$ is in the NVE ensemble using fluctuations in kinetic energy, $K$ \cite{Tuckerman2010}:
\begin{equation}
    \label{eqn:cv}
    C_V = \frac{3 N \kb}{2} \left( 1 - \frac{\langle K^2 \rangle - \langle K \rangle^2}{\frac{3}{2} N \kb^2 \langle T \rangle^2} \right)^{-1}.
\end{equation}

\textcolor{black}{This expression converges the fastest of the terms in Eq. \ref{eqn:heatcapacityrelation}, implying that kinetic energy is very well sampled in the NVE ensemble.} Eqs. \eqref{eqn:heatcapacityrelation}-\eqref{eqn:cv} can be used to calculate C$_P$ with faster convergence than the direct expression in Eq. \eqref{eqn:cp}, under the proviso that the mean density and temperature of the NPT and NVE simulations coincide.

All calculated properties were averaged over 20 different initial conditions created by different seeding of initial particle velocities. 

\section{Results and discussion}

\subsection{Total thermal conductivity}
\label{micro}

\begin{figure}
\begin{center}
{\scalebox{0.6}{\includegraphics{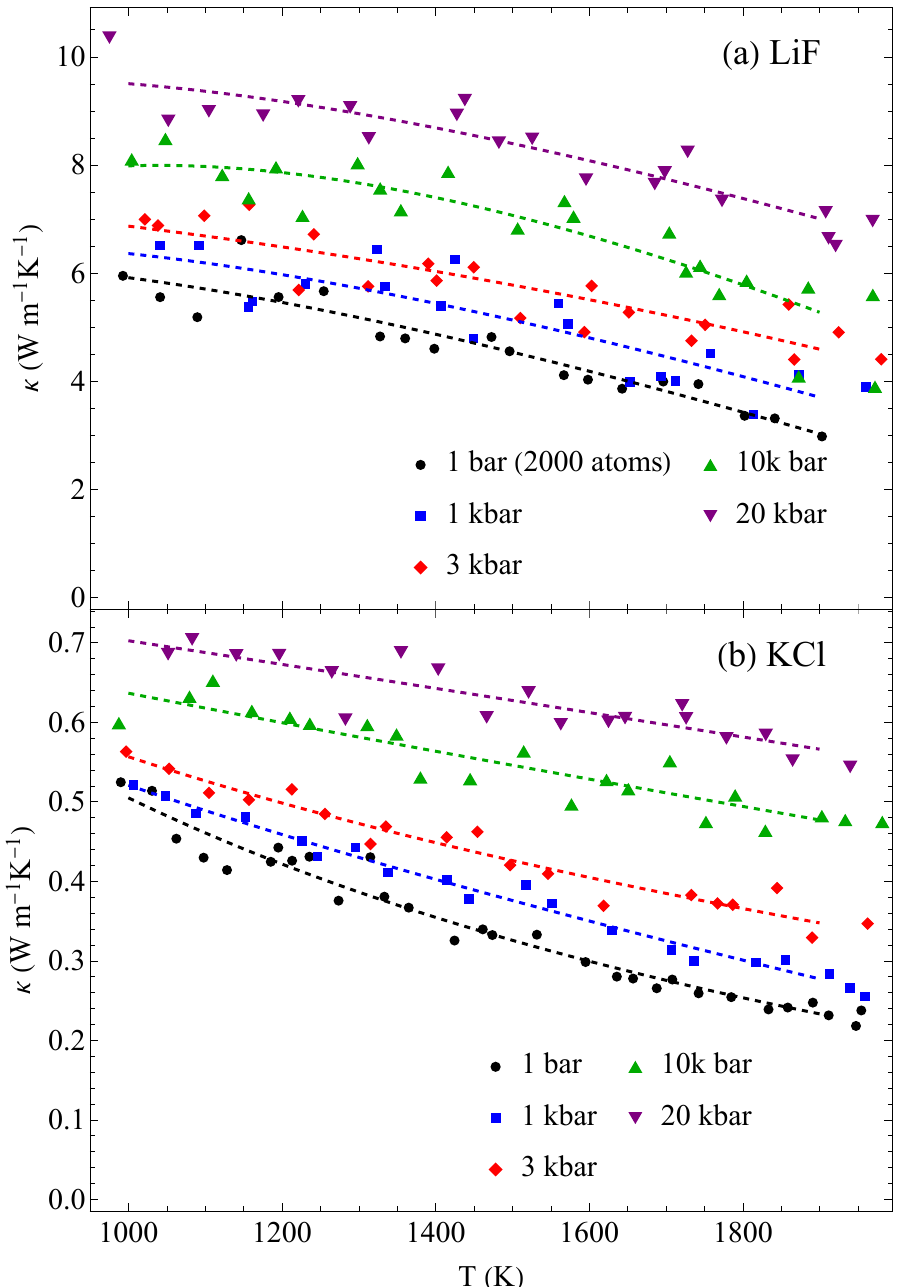}}}
\end{center}
        \caption{(a) Total thermal conductivity, calculated according to Eq. \eqref{eqn:kappa} of LiF shown at multiple pressures. The system size featured was 500 atoms and one 2000 atom system as indicated in legend. 
        (b) Total thermal conductivity of KCl shown at multiple pressures. The system size shown is 1024 atoms.}
        \label{therm_cond}
\end{figure}

We start with the total thermal conductivity, $\kappa$, defined in Eq. \eqref{eqn:kappa}, which directly relates microscopic energy fluctuation and dissipation. It includes the intrinsic conduction plus extrinsic enthalpy advection. We display this thermal conductivity of KCl and LiF in Figure \ref{therm_cond}.

\textcolor{black}{The spread in the data presented here and below is chiefly due to the Green-Kubo integrals of different initial conditions converging on different values, a well known problem \cite{Zhang2015}, with autocorrelation functions involving the heat current being somewhat more pathological than those involving the stress tensor (used to calculate viscosity) in molten salts \cite{Cockrell2025b}.} Figure \ref{therm_cond} shows the expected downward trend of thermal conductivity with increasing temperature. As discussed in the Introduction, this decrease is largely due to the decrease of the speed of sound and the phonon mean free path with temperature. 

We observe that $\kappa$ increases with pressure. High pressure tends to increase liquid relaxation time $\tau$, the average time between atomic jumps. As discussed below in more detail, this increases heat capacity. It also increases the phonon mean free path because phonons can travel for longer before getting scattered. Finally, pressure increases the speed of sound. However, higher pressure also suppresses atomic mobility, which will can be expected to reduce the advective contribution to energy transport (encoded in $L_{q,1}$). As a result of these combined effects, thermal conductivity $\kappa=cvl$ in Eq. \eqref{kappa} increases with pressure as is seen in Figure \ref{therm_cond}.

We now comment on the minima of thermal conductivity discussed in the Introduction. These minima in molecular and noble liquids are due to the crossover between the low-temperature liquidlike and high-temperature gaslike dynamics \cite{prbthermal,brareview,myreview}. In noble and molecular systems, the minima of $\lambda$ are seen in the 100-1000 K range at pressure above the critical point where no boiling transition intervenes at high temperature \cite{prbthermal}. 

Differently from noble and molecular systems, the melting and critical points of molten salts are high. This means that the minima are expected to be seen at very high temperature and above the experimentally available range. Note that if the system boils below the critical point, the smooth minima are substituted by sharp changes (which nevertheless show minima) \cite{sciadv1}. This is expected to be the case for molten salts at operating conditions below the pressure of the critical point. 

This means that although we may not see the minima of thermal conductivity in the temperature range studied, we can interpret the observed decrease of thermal conductivity in Figure 1 in terms of underlying atomistic mechanism. As discussed in the Introduction, this decrease is related to the liquidlike dynamics where each particle oscillates many times around a quasi-equilibrium position before jumping to 
the next one.


\subsection{Intrinsic thermal conductivity}

We now report the thermal conductivity which is accessible to experiments, $\lambda$, as defined in Eq. \eqref{eqn:lambda}. The thermal conductivity $\kappa$ deduced from the fluctuations in microscopic heat current alone overestimate $\lambda$, due to its incorporation of heat advected by partial diffusion currents - the total momentum vanishes but $\mathbf{j}_1$ and $\mathbf{j}_2$ do not. 

\begin{figure}
\begin{center}
{\scalebox{0.65}{\includegraphics{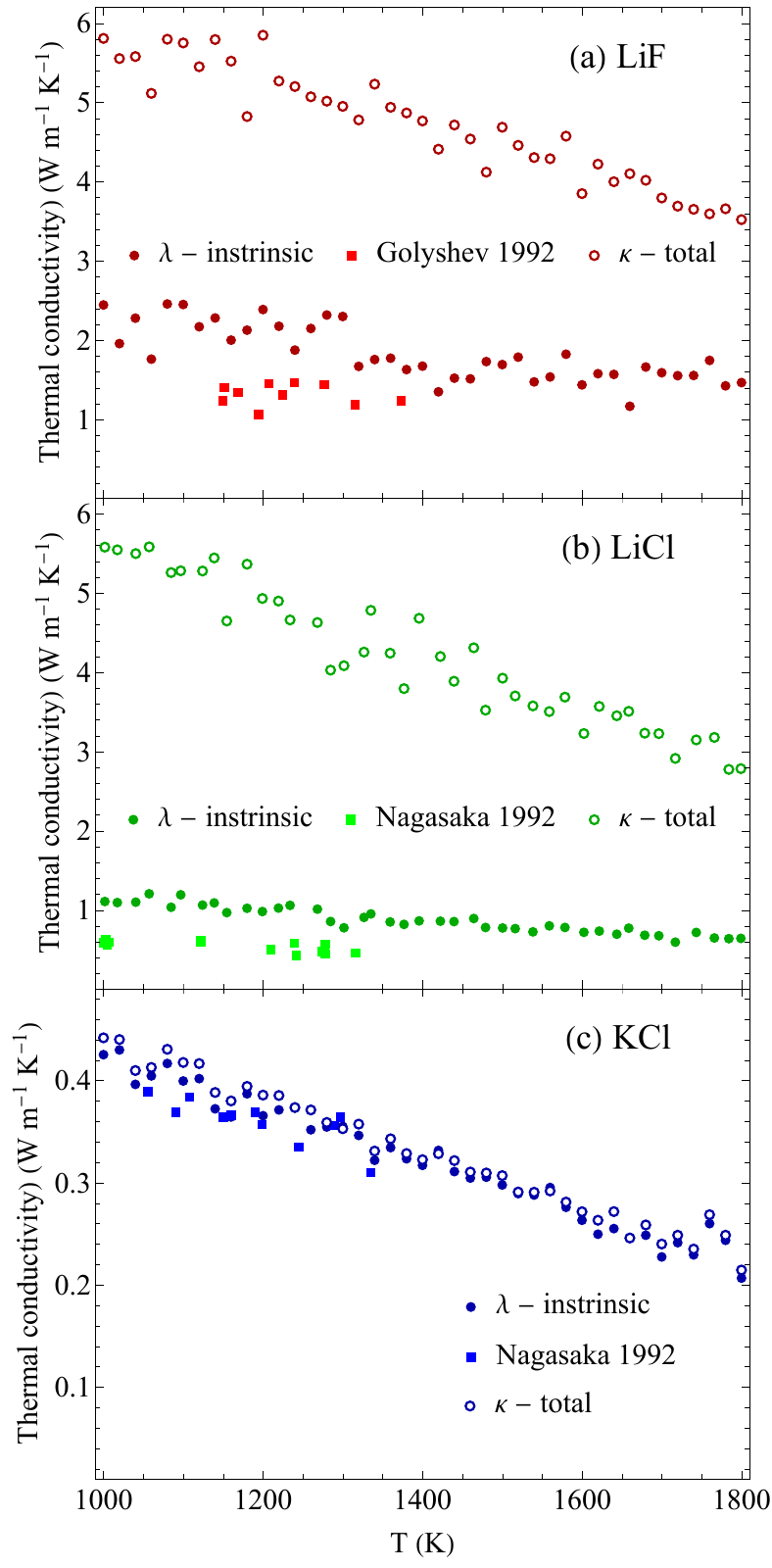}}}
\end{center}
        \caption{Intrinsic thermal conductivity $\lambda$, subtracting mass transfer coupling (according to Eq. \eqref{eqn:lambda}), for (a) LiF; (b) LiCl; (c) KCl \textcolor{black}{each along the 1 bar isobar}, with ``microscopic" $\kappa$ (according to Eq. \eqref{eqn:kappa}) for reference. Experimental intrinsic thermal conductivities from \cite{Golyshev1992} and \cite{Nagasaka1992} are also shown.}
        \label{fig:macrolambda}
\end{figure}

To see this effect clearly, we have increased the number of calculated temperature data points and improved statistics for one pressure point of 1 bar corresponding to experimental conditions. We show the calculated thermal conductivities $\kappa$ and the $\lambda$ for LiF, LiCl, and KCl in Fig. \ref{fig:macrolambda}a-b. \textcolor{black}{We report similar results to those shown at fewer temperatures in previous studies \cite{Ohtori2009,Galamba2007} - we note that, though these studies choose charge current density rather than momentum density as a variable coupled to the heat current density, in these models all motion of charge is from direct advection by the ions, meaning that charge and mass transport are entirely interdependent.}

To compare our simulation results to experiments, we use the experimental thermal conductivity data recommended in Ref. \cite{gheribi}. These include the data from Refs. \cite{Golyshev1992} and \cite{Nagasaka1992}. These data are overlaid on Fig. \ref{fig:macrolambda}. Experimental measurements of thermal conductivity are challenging, and the data for LiF differ by a factor 2, consistent with trends reviewed in Ref. \cite{zhao-review}. Nonetheless, the calculated intrinsic thermal conductivity lie closely to previous data for LiF and KCl, with qualitative trends of all experimental data matching those of our data.

The calculated intrinsic thermal conductivity is closer to the experimental data as compared to the total thermal conductivity in Figure \ref{therm_cond}, with this effect more pronounced in LiF and LiCl than KCl. In LiF and LiCl, the presence of the light species Li significantly enhances the transport of energy, such that $\kappa$ exceeds $\lambda$ by a factor of 2-3. \textcolor{black}{In KCl, where masses are nearly identical, there is no significant difference between $\lambda$ and $\kappa$. This is because the cross-correlation term, $L_{q,1}^2$ is much smaller than the mass transport coefficient $L_{1,1}$, such that the term subtracted from $L_{q,q}$ is negligible, In other words, the heat advected by mass diffusion is negligible compared to direct conduction, and mass coupling effects may therefore be ignored. Indeed, K and Cl will have nearly identical velocity distributions such that one partial current $\mathbf{j}_1$ does not advect significantly more energy, on average, than its pair, $\mathbf{j}_2 = - \mathbf{j}_1$. Inspecting Eq. \ref{eqn:qcurrent} we see that,} assuming roughly comparable forces, both the kinetic and potential energy transport are significantly enhanced by the larger velocity of Li compared to F or Cl, with $\sqrt{m_2/m_1} \approx 1.65$.

The same trend of thermal conductivity decreasing with temperature is seen in $\lambda$ and $\kappa$ of both liquids, while the quantity $L_{1,q}^2/L_{1,1}$ appearing in Eq. \eqref{eqn:lambda} remains roughly constant, implying that the explanation for the decrease of $\kappa$ in terms of a decreasing phonon mean free path and speed of sound remains qualitatively correct. Indeed, mass transfer (which takes place \textit{via} atomic jumps) is a contributing mechanism to phonon scattering.

\begin{figure}
\begin{center}
{\scalebox{0.65}{\includegraphics{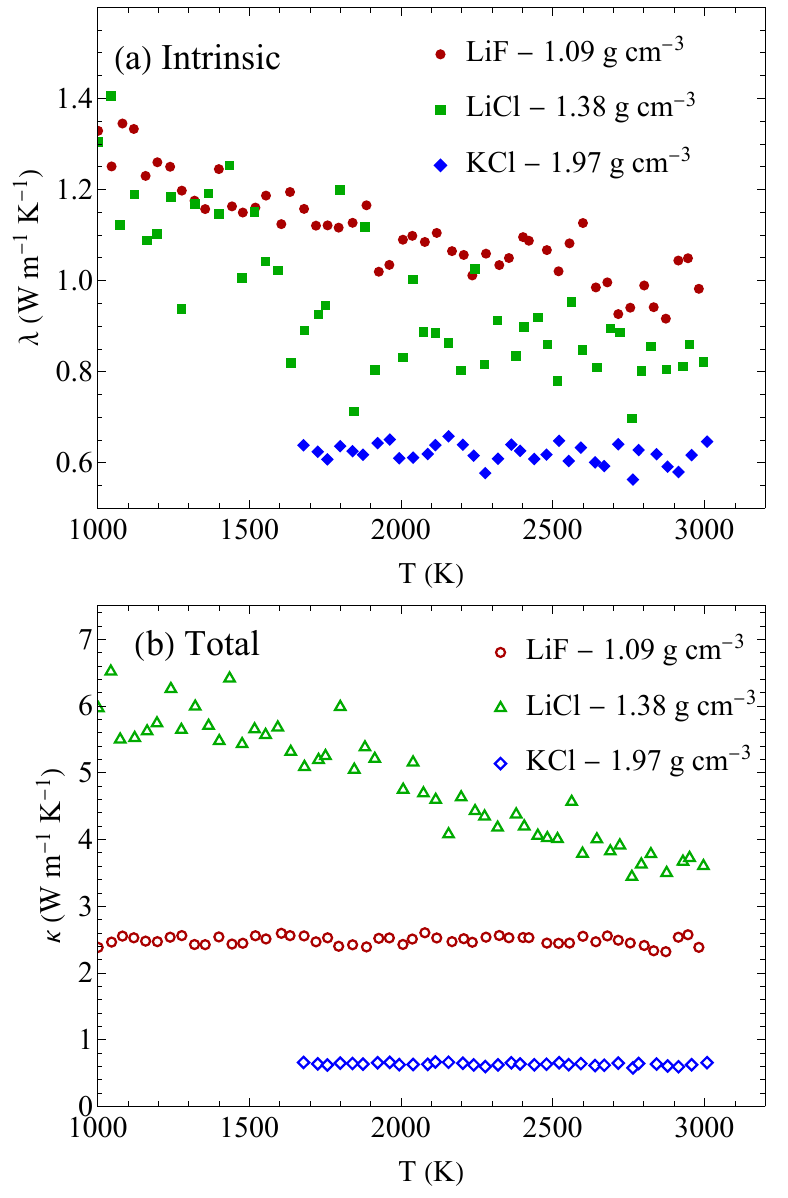}}}
\end{center}
        \caption{(a) Intrinsic thermal conductivity $\lambda$; (b) total thermal conductivity $\kappa$, of LiF, LiCl, and KCl as a function of temperature along isochores.}
        \label{fig:tcdensity}
\end{figure}

The thermal conductivity of pure molten alkali halides has previously been suggested to depend mostly on density, and on temperature only \textit{via} density \cite{Ohtori2009,Ishii2014}. We investigate this in detail by simulating an isochore for each composition, each at many temperatures in order to clearly see the evolution at constant density. The total and intrinsic thermal conductivities along each isochore are displayed in Fig. \ref{fig:tcdensity}. Here we see very clearly a strong temperature dependence, independent of density, of the intrinsic thermal conductivity in the Li-containing melts, and of the total thermal conductivity in LiCl. Meanwhile the intrinsic thermal conductivity of KCl and the total thermal conductivity of both LiF and KCl exhibit no trend with temperature, indicating indeed that the density dependence is the predominant factor in determining the conductivity, at least in KCl. In the case of LiF, these data imply that the intrinsic heat conduction and heat advection have competing temperature responses, with the former decreasing with increasing temperature at roughly the same rate as the latter increases.

\subsection{Heat capacity}

Calculating thermal diffusivity involves Eq. \eqref{eqn:thdif}. The challenging property to calculate is the constant-pressure heat capacity which is subject to considerable fluctuations at high temperature, even with very long MD runs and averaging over many initial conditions. As mentioned above, calculating $C_P$ from Eq. \eqref{eqn:heatcapacityrelation} involving constant-volume heat capacity $C_V$ reduces fluctuations. Discussing $C_V$ is also interesting on its own because it is related to the system energy and the number of degrees of freedom in the system (as opposed to $C_P$, which additionally depends on thermal expansion and bulk modulus).

\begin{figure}
\begin{center}
{\scalebox{0.55}{\includegraphics{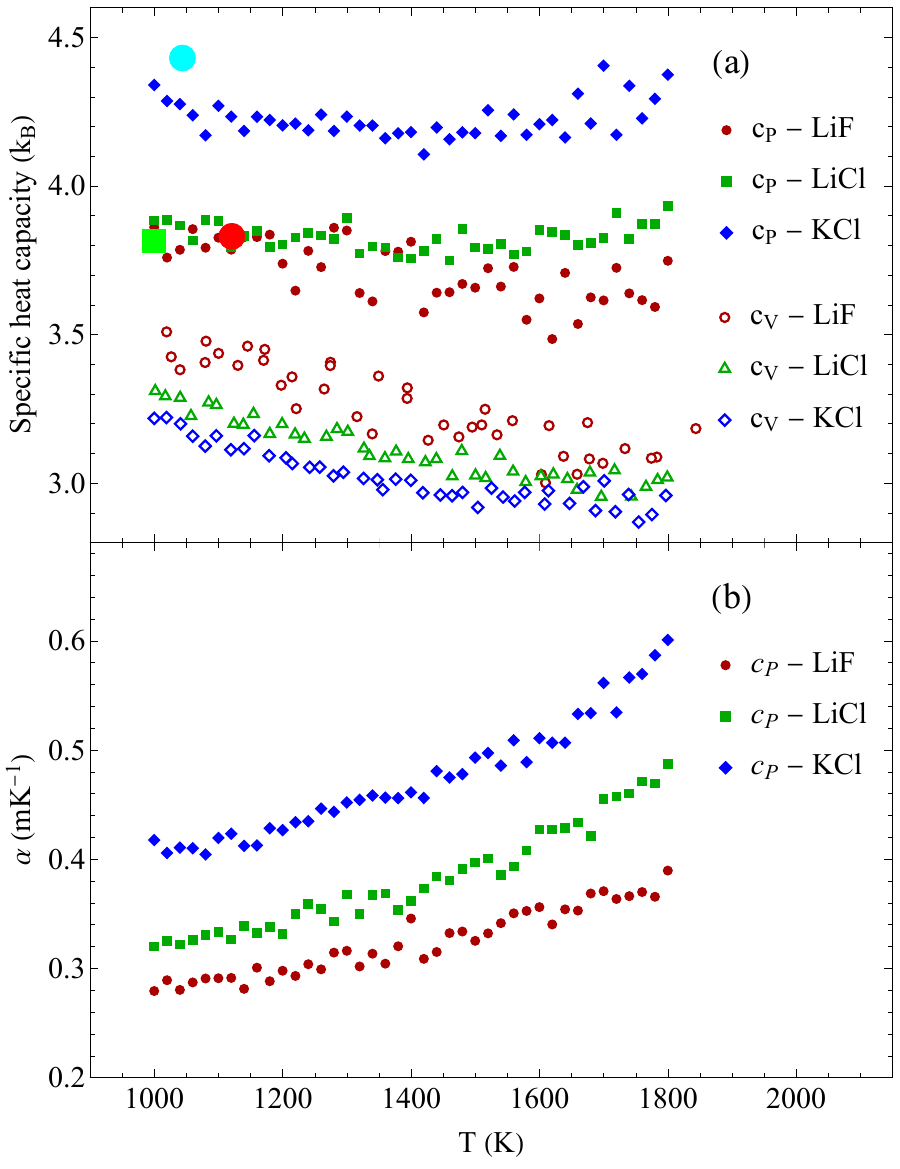}}}
\end{center}
        \caption{(a) Constant-pressure and constant-volume specific heats (heat capacities per atom) of LiF, LiCl, and KCl. Also shown in larger solid symbols are experimental constant-pressure heat capacities near the melting line from NIST \cite{nist}; (b) thermal expansion coefficients of LiF, LiCl, and KCl.}
        \label{heat_cap_p}
\end{figure}

Figure \ref{heat_cap_p}a shows the calculated specific (per particle) isobaric and isochoric heat capacities, $c_P$ and $c_V$. The decrease of $c_V$ with increasing temperature follows the common trend of all liquids \cite{ropp,mybook,proctor1,proctor2,chen-review,withbook,nist} and is due to the loss of propagating transverse phonons below the hopping frequency $\omega_{\rm F}=\frac{1}{\tau}$, where $\tau$ is liquid relaxation time, time between two consecutive particle jumps in the liquid. Quantitatively, liquid $c_V$ is described by the phonon theory of liquid thermodynamics \cite{ropp,mybook,proctor1,proctor2,chen-review,withbook} as follows.

The theory is based on the main premise of statistical physics: properties of an interacting system are governed by its excitations \cite{landaustat,landaustat1}. In solids, these are collective excitations, phonons. The theory further focuses on energy $E$ as the primary property in statistical physics \cite{landaustat}. The energy of phonon excitations in the liquid is the sum of the longitudinal phonon energy and the energy of gapped transverse phonons with frequency $\omega>\omega_{\rm F}$. Adding the energy of these excitations to the energy of diffusing atoms gives the liquid energy as \cite{ropp,mybook,proctor1,proctor2,chen-review,withbook}

\begin{equation}
E=N T\left(3-\left(\frac{\omega_{\rm F}}{\omega_{\rm D}}\right)^3\right)
\label{harmo}
\end{equation}

\noindent where $\omega_{\rm D}$ is the maximal Debye frequency that sets the high-temperature limit for $\omega_{\rm F}$ and the energy of each phonon is $T$ in the classical and harmonic case. Here and below, $k_{\rm B}=1$. The hopping frequency $\omega_{\rm F}$ increases with temperature. This gives the reduction of the energy slope and the decrease of $c_V$ with increasing temperature in Figure \ref{heat_cap_p}. \textcolor{black}{This equation accounts for the entire liquid energy, using the harmonic approximation to relate the diffusive energy to the vibrational energy \cite{ropp}. Though the harmonic approximation is seemingly inapplicable to liquid vibrations, this model has been independently tested in liquids with a variety of bonding types \cite{proctor1} - the model, though falsifiable, was not falsified. The parameter $\omega_{\rm D}$, though inspired by the Debye theory of the solid state, here represents roughly the maximal frequency of atomic oscillation, and takes physically sensible values of around $10^{13}\  \mathrm{s}^{-1}$ when Eq. \ref{harmo} is fit to experimental data. The representation of liquid energy as a gapped spectrum of harmonic collective excitation therefore appears to be roughly physically correct in a statistical sense \cite{proctor1}, even though individual particle motion is not so easily stratified into oscillation and diffusion.}

At low temperature, Eq. \eqref{harmo} predicts $c_V\approx 3$ when $\omega_{\rm}\ll\omega_{\rm D}$. This is the solid Dulong-Petit result. $c_V$ in Figure \ref{heat_cap_p} is slightly larger than $3$ at low temperature, consistent with experimental $c_V$ in many noble, molecular and metallic liquids \cite{ropp,mybook} (as well as solids). This increase above $c_V=3$ is due to the anharmonic effects increasing the energy of each phonon above the harmonic value of $T$ (recall that this harmonic value is assumed in Eq. \eqref{harmo}). Accounting for this anharmonicity describes the experimental $c_V$ and its increase above $3$ at low temperature \cite{ropp,mybook,proctor1,proctor2}.

The decrease of $c_V$ with \textcolor{black}{increasing} temperature in liquids represents a general effect specific to the liquid state of matter. In solids, the phase space available to phonons is always fixed: at any temperature and pressure, this phase space is set by $3N$ propagating phonons, where $N$ is the number of atoms. On the other hand, the phase space available to phonons in liquids is {\it variable} and reduces with increasing temperature \cite{mybook}.

Temperature dependence of constant-pressure heat capacity $c_P$ in Eq. \eqref{eqn:heatcapacityrelation} has two competing terms: the decrease of $c_V$ with temperature in the first term and the increase due to the second $\propto T$ term (additionally, $V$ increases with temperature, as does $\alpha$). As a result, $c_P$ is predicted to show a weaker decrease with temperature. Consistent with this prediction, $c_P$ shows weaker temperature dependence in LiF. In KCl, $c_P$ is nearly constant in almost entire temperature range and starts to increase in the high-temperature range due to the second term in Eq. \eqref{eqn:heatcapacityrelation}. Experimental enthalpies close to the melting point predict $c_P$ to be 3.83$\kb$ and 4.43$\kb$ for LiF and KCl respectively \cite{nist}, demonstrating excellent agreement with our modelling results.

We note $c_V$ of the molten salts increases with the decreasing mass of the constituent atoms - KCl has the smallest, followed by LiCl and LiF. Meanwhile the opposite trend is seen in $c_P$. This is principally due to the thermal expansion coefficient, $\av$, displayed in Fig. \ref{heat_cap_p}b, which is far larger in the compounds with heavier ions. The heat capacity $c_P$, and therefore thermal diffusivity $\ath$, of pure molten salts are therefore increased and decreased respectively due to the increased thermal expansion from heavier ions, as discussed further below.

We therefore find that discussing heat capacity in molten salts is important from at least two perspectives. First, it is related to key thermodynamic properties and their understanding on the basis of excitations in the system, the primary goal of statistical physics \cite{landaustat1}. Second, these phonon excitations carry heat in liquids and set liquid heat capacity, directly affecting thermal conductivity in Eq. \eqref{kappa}.

\subsection{Thermal diffusivity}

\begin{figure}
\begin{center}
{\scalebox{0.4}{\includegraphics{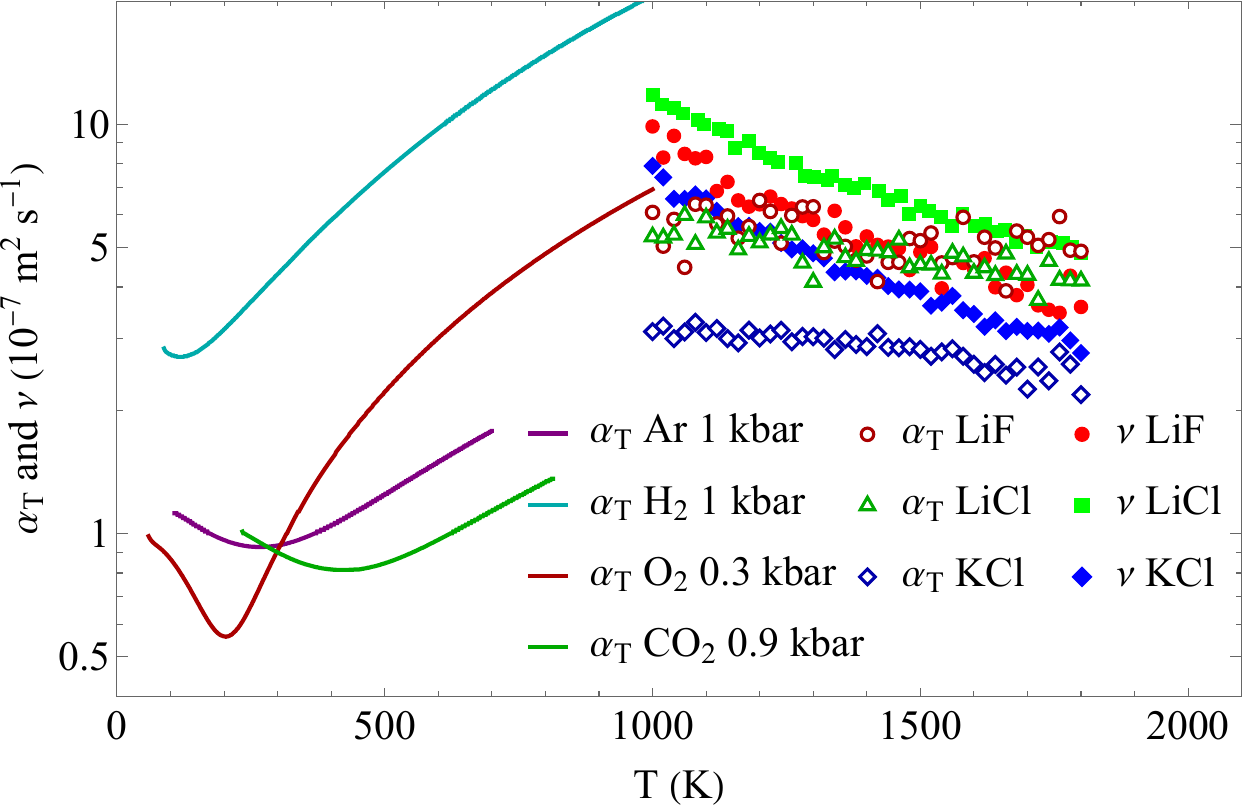}}}
\end{center}
        \caption{Thermal diffusivity and kinematic viscosity of pure salts from MD simulations, each at 1 bar. For comparison, we show experimental thermal diffusivity of Ar, H$_2$, O$_2$ and CO$_2$ \cite{nist}.}
        \label{minimum_compare}
\end{figure}

Figure \ref{minimum_compare} shows the calculated thermal diffusivity of LiF, LiCl, and KCl. In these compounds $\alpha_T$ is on the order of $10^{-7}\frac {{\rm m}^2}{\rm s}$ and, close to the theoretical bound \eqref{alphamin} or \eqref{alphafund}. The intrinsic thermal conductivity, $\lambda$, is largest for the lightest ions (LiF), and decreases as the ion mass increases. The isobaric heat capacity follows the opposite trend, however we here see that the thermal diffusivity of LiF and LiCl are very similar. This is due to the appearance of $\rho c_P$ in the denominator of Eq. \ref{eqn:thdif}. This quantity is the isobaric heat capacity per unit volume, and though $c_P$ (per ion) for LiF and LiCl are also very similar, the ionic density of LiF compared to LiCl (roughly a factor of 2, independent of temperature) decreases this volumetric heat capacity. The discrepancy in $\lambda$ between LiF and LiCl is therefore eliminated in $\ath$ by their ionic density differences. KCl, meanwhile, has a larger isobaric specific heat capacity and lower thermal conductivity than both, greatly bringing down $\ath$, with its lower ionic density bringing it up slightly as each unit volume of space carries less heat.

It should be noted that the fundamental lower limits of thermal diffusivity in Eqs. \eqref{alphamin} and \eqref{alphafund} reflect the ``microscopic'' mechanism involving the crossover between the low-temperature liquidlike and high-temperature gaslike dynamics and do not take the effect of mass current operating in binary systems into account. This applies trivially to molecular and noble liquids where this effect does not operate \cite{prbthermal,brareview,myreview}. Nevertheless, we observe in Figures \ref{fig:macrolambda} and \ref{minimum_compare} that the intrinsic thermal conductivity respects the bound, and therefore that the total conductivity, increased by the mass current effect, likewise will not violate the bound. Molten salts and ionic liquids therefore join the family of other types of liquids where this was seen on the basis of experimental data \cite{prbthermal,brareview,myreview}.

We also note that the smallest values of $\alpha_T$ seen in there compounds are close to the minimum of thermal diffusivity in some molecular liquids such as CO$_2$ but are higher than in other molecular and noble liquids. One reason for this difference is discussed in section \ref{micro}: differently from noble and molecular liquids, the thermal conductivity in molten salts may not be reached because (a) their melting and boiling points are high and (b) the boiling transition intervenes below the critical point studied here. Another reason is related to the mass effect: the theoretical minima \eqref{alphamin} vary, albeit weakly, with the molecular mass. This contributes to the difference between the theoretical and experimental minima, which for different liquid can vary by a factor in the range 1-4 \cite{prbthermal}.

We note that throughout this paper, we have discussed thermal conductivity and thermal diffusivity due to lattice effects where heat carriers are either phonons or atoms. In metals, including metallic liquids, the dominant heat careers are electrons. The lower bound of thermal diffusivity in metallic liquids is due to electronic contribution and is set by \cite{brareview,myreview} 

\begin{equation}
    \alpha^e_{\mathrm{min}} = \frac{\hbar}{m_e}\approx 10^{-4}~\frac{{\rm m^2}}{\rm s}
\end{equation}

This bound is 3 orders of magnitude larger than the bound Eq. \eqref{alphamin} due to the difference between the proton and electron masses and reflects the difference between phonons and electrons as carriers of heat \cite{brareview,myreview}.

\subsection{Comparison of thermal diffusivity and kinematic viscosity}

It is interesting to compare thermal diffusivity, the measure of thermal energy diffusion to another liquid property, kinematic viscosity, which is the measure of momentum diffusion. Despite different physical mechanisms involved in setting the two properties, it was found that experimental thermal diffusivity and kinematic viscosity  (a) lie close to each other and (b) show minima which are also close to each other \cite{prbthermal}. This has been shown for many molecular, noble and hydrogen-bonded liquids. However, this has not been tested in ionic liquids. 

We have earlier calculated kinematic viscosity $\nu$ of LiF using molecular dynamics simulations \cite{withington2024}. Here we also calculated $\nu$ for LiCl and KCl, and show $\nu$ for all systems in Figure 3. We observe that kinematic viscosities are not far from thermal diffusivities of both molten salts. Similarly to thermal diffusivity, viscosity decreases with temperature in both systems. This is consistent with the liquidlike regime of dynamics where each particle oscillates many times before jumping to the next position (note that viscosity increases with temperature under gaslike dynamics). 

The closeness of similar temperature behavior of thermal diffusivity and kinematic viscosity is consistent with the similarities of both properties in noble, molecular and hydrogen-bonded liquids \cite{prbthermal}. This similarity gives an interesting comparison between two different transport properties of molten salts. \textcolor{black}{While thermal conductivity quantifies the rate of heat transport through a system, thermal diffusivity quantifies the rate of \textit{temperature} transport - likewise the kinematic viscosity describes \textit{velocity} transport rather than momentum transport. Each has the same units as the diffusion coefficient and ``normalises" for system-specific effects density or heat capacity (also known as thermal mass) in order to describe the diffusion through time and space of these fields, velocity and temperature, directly. The similarity of $\alpha_{T}$ and $\nu$ (and the equality of their predicted fundamental bounds) in molten salts implies that these liquids respond to these thermodynamic forces \textit{via} similar atomic mechanisms - phonons operating in a gapped momentum state \cite{ropp,Cockrell2025b}.}

\section{Summary}

We calculated key thermal properties of molten salts, including thermal conductivity, thermal diffusivity, and heat capacity. Combining the simulated results with recent theoretical insights, we provided atomistic interpretation of these properties at the level of atomistic dynamics and phonon excitations in liquids. Separate definitions of thermal conductivity, which differ by the implicit inclusion or explicit exclusion of energy transport \textit{via} mass diffusion, do not qualitatively differ, demonstrating the ability of the phonon model to qualitatively and generally predict energy transfer in binary ionic liquids.  We also found thermal diffusivity of molten salts is close to another of their transport properties, kinematic viscosity, and discussed the origin of this effect. 

Given the scarcity of data related to molten salts and their general understanding \cite{zhao-review}, our discussion will be useful in recent and emergent applications of these systems. It also adds to general understanding of liquid properties. This is related to a more fundamental research which is important in view that the liquid state has traditionally been understood far less as compared to solids and gases \cite{ropp,mybook}.

We are grateful to EPSRC (grants No. EP/X011607/1 and EP/W029006/1) and Queen Mary University of London for support. For all  simulations conducted, this research utilised Queen Mary's Apocrita HPC facility, supported by QMUL Research-IT. http://doi.org/10.5281/zenodo.438045.
The US researchers acknowledge financial support from the Department of Energy, Office of Nuclear Energy’s Nuclear Energy University Programs via Award No. DE-NE0009288.


\bibliography{Z-Biblography}

\end{document}